\documentstyle[12pt,epsf]{article}

\begin{document}

%----------------Шапка----------------------------------------------

\begin{center}

%\vskip 7cm 
{\tt {\Large \bf
\vskip 7cm 
\mbox{Is the `Soft Pomeron' Valid for the }
\vskip 0.5cm 
\mbox{Description of the Data from HERA?}
}}
\vskip 3cm 

\mbox{V.A.Petrov and A.V.Prokudin}
 
\mbox{Division of Theoretical Physics, IHEP,}

\mbox{{\it 142 284} Protvino, Russia}

 \vskip 1.75cm 
{\bf 
\mbox{Abstract}}

\begin{flushleft}
It is established in the paper that experimental data on deeply virtual exclusive electroproduction of
$\rho^0$- and $\phi$-mesons may be  described fairly well in the framework of a generalized Regge approach
without additional, $Q^2$-dependent, singularities in the $J$-plane.
\end{flushleft} 
\end{center}

\thispagestyle{empty} 
\setcounter{page}{0} 
\newpage
%---------------Содержание------------------------------------------

%----------------Ведение---------------------------------------------

\section*{Introduction}

Experiments at $ep$-collider HERA have revealed that the cross section
for exclusive production of light mesons by the virtual photon grows with
increasing energy faster than the cross section for production of 
the same mesons by the real photon \cite{1}. Besides, the cross section for production
of heavy mesons ($J/\Psi$ etc)by the real photon also grows with energy faster than
the cross-section for production of light mesons ($\rho ,\omega ,\phi$).

Therefore, with the presence of the second energy scale (besides the collision energy)
which is virtuality of the photon $Q^2$ or the mass of the produced vector 
meson, the dependence on energy gets stronger.

There are several opinions on this interesting effect, and the most popular
and wide spread (it seems to be accepted by experimentalists) is that
besides the usual, considered the leading (i.e. the most right-hand) singularity
in the $J$-plane and called {\bf {\it "pomeron"\footnote{
Historically, pomeron is a trajectory of the pole with even signature
and quantum numbers of the vacuum, and $\alpha (0)=1$.
At present it is believed that $\alpha (0) >1$. }}}
there exists another singularity, which is located to the right of the former and thus 
the "pomeron" is not the leading singularity any longer.
The peculiarity of the latter singularity (which is called "hard pomeron" sometimes)
is that it exists only with the presence of "hardness" in the proccess (i.e. either
high virtuality or heavy meson mass, like $J/\Psi$).
The "hard pomeron" is considered to play the main role in the effect.
The hypothesis seems to be rather convincing when we look at fig.1 \cite{1} and fig.2 \cite{9},
where the experimental data and the dependence on energy of the cross-section
in the framework of the "hard pomeron" dominance are represented.

Nevertheless, it leads us to some serious questions about consistency
with general principles of the theory, like, for instance, unitarity 
or others less common but quite proved ideas like "maximal analyticty of the 2nd type" \cite
{2}.

%----------------Картинка эксп.--------------------------------------

%\begin{figure}
%\TrimTop{2cm}
%\ForceHeight{14cm}
%\ForceWidth{0.8\textwidth}
%\BoxedEPSF{exper.eps}
%\label{111}
%\caption{ The $W(GeV)$-dependence of the cross-section.}
%\end{figure}
Let's consider one of the processes being explored at HERA 
$$
\gamma^* + p \rightarrow \rho_0 +p ,%
$$
and suppose that the energy dependence of the cross section
$$
\sigma_{\gamma^* p \rightarrow \rho_0 p} \sim \int d(\mbox{phase space})\mid
T_{\gamma^* p \rightarrow \rho_0 p} \mid^2 ,%
$$
is determined by the "hard pomeron"

%----------------Картинка эксп.--------------------------------------

\begin{equation}
\label{*} T_{\gamma^* p \rightarrow \rho_0 p} \sim s^{1+\lambda} 
\mbox{ , where    }
\lambda \simeq 0.2 
\mbox{ (phenomenological value). } 
\end{equation}
The following inequality takes place 
$$
\mid Im T_{\gamma^* p \rightarrow \rho_0 p} \mid^2 \leq \mid Im T_{\gamma^*
p \rightarrow \gamma^* p} \mid \mid Im T_{\rho_0 p \rightarrow \rho_0 p}
\mid 
$$
According to (~\ref{*}) the left part grows as $s^{2(1+\lambda)}$
the right as $s^{2+\lambda+\Delta}$, where $\Delta = \alpha(0)-1$ is connected with
the "soft" pomeron which controls pure hadronic process $\rho_0+p \rightarrow \rho_0+p$.
So we obtain
$$
\lambda \leq \Delta,%
$$
with full contradiction with the hypothesis of the "hard" pomeron dominance.
%---------------------exper fig. on rho0------------
\begin{figure}[t]
\label{111}
\vskip  -3cm
\hskip 2cm \vbox to 10cm {\hbox to 10cm{\epsfxsize=10cm\epsffile{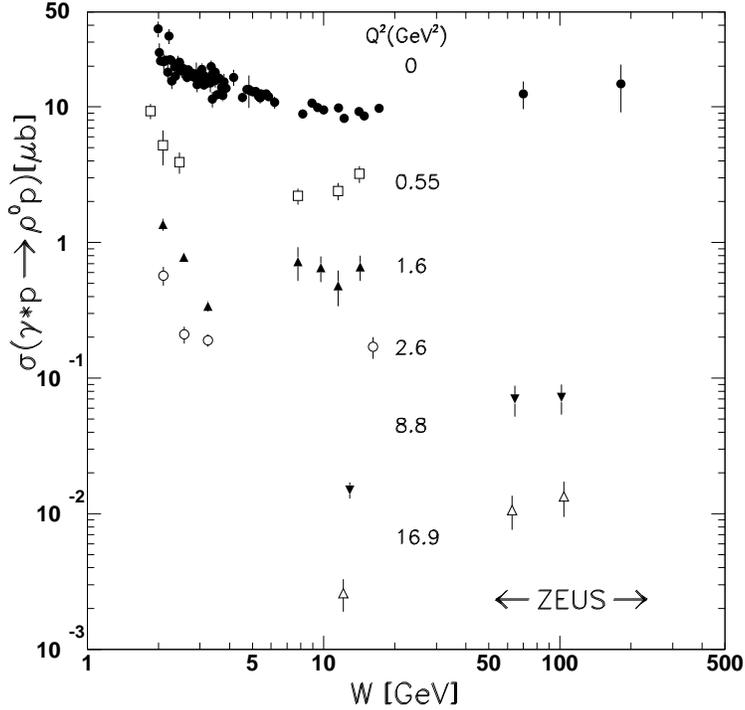}}}
\vskip  2cm
\caption{ The $W$-dependence of the cross-section.}
\end{figure}
%----------------------------------------------------

One could think, that "unitarization" of the powerlike (Born) behaviour
would help us to avoid this conclusion, but it seems not to take place and
the "paradox" remains \cite{3}. So, it looks not to be with no purpose to search
for new explanation of the effect.
For example, in \cite{4} an alternative idea was proposed  for description
of the total cross-section. The idea is that the onset of
"preliminary \footnote{"Preliminary" in the sence that in the region of extra-high
energy significant modifications are possible.}
asymptotics", which is reached in "soft" processes, only starts
in "hard" processes. In other words, the threshold of "preliminary
asymptotics" is increased due to "hardness". Unfortunately, in the paper
 \cite{4} authors gave just a phenomenological description, which
drops out of the pure Regge framework. The description is abound
in peculiar parameterizations of residues, and tricky functions,
and a great number of parameters. It makes the argument not
quite convincing.
 Preasymptotical character of the energy dependence of structure
functions in the energy region of HERA was also disputed in \cite{5}.

%\begin{figure}
%\TrimTop{2cm}
%%\vbox to 23cm {\hbox to 14cm{\epsfxsize=14cm\epsffile{tex6.ps}}}
%\ForceHeight{14cm}
%\ForceWidth{0.8\textwidth}
%\BoxedEPSF{experp.eps}
%\label{222} 
%\caption{ The $W(GeV)$-dependence of the cross-section.}
%\end{figure}
In this paper we suggest a concrete method based on using Regge
pole conception and the scenario of delaying asymptotics is
realized in its framework. We formulate the problem in a
different way: If it's possible to describe the data in
the Regge-eikonal framework, and if 'yes', then what is the 
$Q^2$-dependence of residues etc due to the data?

The positive answer to the first question raises the problem of the  "hard pomeron" status.
 The result of extraction of the $Q^2$-dependence of the residues
 puts in fact the problem of the theoretical
investigation of transition form-factors (reggeon-vector meson).
%---------------Основная часть-------------------------

\section*{Regge-eikonal model for processes with virtual particles.}
The amplitude of the process 
$$
\gamma^* + N \rightarrow V + N\mbox{ ,  }  
$$
$T_{\gamma^* N \rightarrow V N}(\vec b;s,Q^2)$,
has the following form in the impact parameter representation
(* always means "off-mass-shell")

\begin{equation}
\label{1}T_{\gamma ^{*}N\rightarrow VN}=\sum_{V^{\prime }}c_{V^{\prime
}}(Q^2)\frac{\delta _{V^{\prime }V^{\prime }}^{*}}{\delta _{V^{\prime
}V^{\prime }}}T_{V^{\prime }N\rightarrow VN}+\sum_{V^{\prime }\neq
V}c_{V^{\prime }}(Q^2)(\delta _{V^{\prime }V}^{*}-\frac{\delta
_{V^{\prime }V^{\prime }}^{*}}{\delta _{V^{\prime }V^{\prime}}}\delta
_{V^{\prime }V})e^{i\delta _{VV}} 
\end{equation}
where $c_V(q^2)=\frac{\mu^2/f_V}{\mu^2+Q^2}$, $f_V$ the coupling constant of the meson $V$ to the electromagnetic
current,
$\mu$ the mass of the vector meson,
$\delta_{AB}(\vec b; s, Q^2)$  generalization of the Born amplitude
for nondiagonal (and off-mass-shell) process
$$
A+N \rightarrow B+N 
$$

%---------------------exper fig. on phi------------
\begin{figure}[t]
\label{222}
\vskip  -3cm
\hskip 2cm \vbox to 10cm {\hbox to 10cm{\epsfxsize=10cm\epsffile{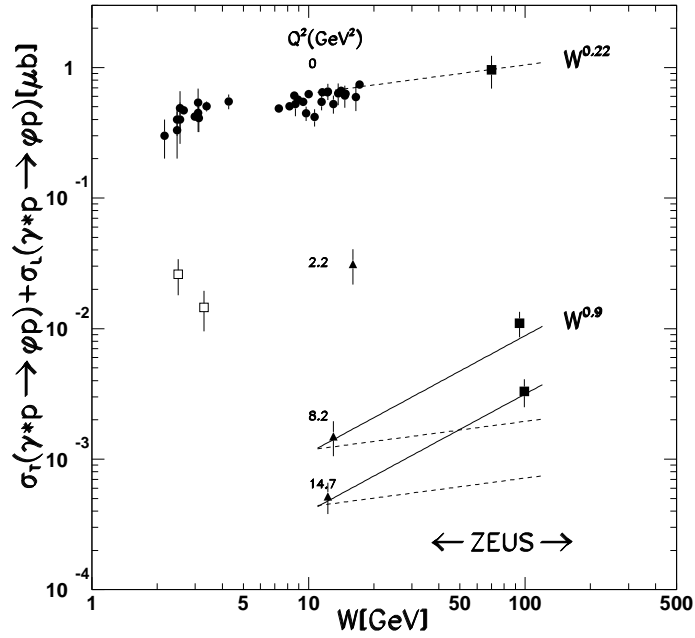}}}
%\vskip  2cm
\caption{ The $W$-dependence of the cross-section.}
\end{figure}
%----------------------------------------------------

In the Regge-eikonal framework $\delta _{AB}(\vec b;s,Q^2)$ has form

$$
\delta _{AB}(\overrightarrow{b};s,Q^2)=\int \frac{dt}{16\pi s}J_0(b\sqrt{-t})\sum_{n} 
g_{AB}^n(Q^2,t)g_{NN}^n(t)\xi _n(t)\left( \frac
{s}{\mu ^2+Q^2}\right)^{\alpha _n (t)} 
$$
where sum goes in all relevant reggeons
(with trajectories $\alpha_n (t)$ and signature factors $\xi_n(t)$).

These formulas are effective scalar,
because we do not take into account different polarizations of
vector mesons. In this sence the method used here is valid  for processes
like $W^{*}(Z^{*})+N\rightarrow \pi +N$. 

For further analysis we retain in (\ref{1})  only
contributions of the pomeron and the secondary reggeon $\alpha _R(t)$ in
the linear approximation in $t$ with parameters: 
$$
\begin{array}{c}
\Delta \equiv \alpha _P-1=0.075 \\ 
\alpha _P^{\prime }(0)=0.25
\end{array}
$$
$$
\begin{array}{c}
\eta \equiv 1-\alpha _R=0.46 \\ 
\alpha _R^{\prime }(0)=1.00
\end{array}
$$
partially taken from papers \cite{7}, and tested for the total and
differential cross-section description.

Formula (\ref{1}) is adjusted with unitarity, which, generally,
 in the case of virtual particles does not lead to the Martin-Froissart
bound \cite{8} for asymptotic behaviour  of the cross-section,
and even allows for such a powerlike growth as $s^\Delta$ \cite{6}.
Taking it into consideration we will investigate the first Born term only
with following estimation of "unitarity corrections".
$$
T_{\gamma ^{*}N\rightarrow VN}=\delta _{\gamma ^{*}V}[1+i\delta _{VV}(s,b)]+%
 i\sum_{V^{\prime }\neq V}c_{V^{\prime }}(Q^2)\delta
_{V^{^{*\prime }}V^{^{\prime }}}\delta _{V^{^{\prime }}V},
$$
where $ \delta _{\gamma ^{*}V}=\sum_{V^{^{\prime }}}c_{V^{\prime
}}(Q^2)\delta _{V^{^{*\prime }}V}(s,b,Q^2) $.
The analysis shows that the first correction contribution is
less than $10\%$ (the second $0.1\%$), so with present 
experimental accuracy we decided to consider Born term only.

Then one has
$$
\delta _{\gamma ^{*}V}(s,t,Q^2)=
c_P(Q^2)e^{\frac{1}{4}R_{*P}^2\cdot t}\xi _P(t)\left( \frac {s}{Q^2+\mu
^2}\right) ^{\alpha _P(t)}+ \\
c_R(Q^2)e^{\frac{1}{4}R_{*R}^2\cdot t}\xi _R(t)\left( \frac {s}{Q^2+\mu
^2}\right) ^{\alpha _R(t)}
$$
Let's transform the scattering amplitude into
 the impact parameter representation
$$
T_{\gamma ^{*}N\rightarrow VN}\simeq \delta _{\gamma
^* V}(s,b,Q^2)= 
\frac{c_P(Q^2)\xi _P(0)}{\pi (Q^2+\mu ^2)\widetilde{R}_{*P}^2} \left( \frac
{s}{Q^2+\mu ^2}\right) ^\Delta
e^{-\frac{b^2}{\widetilde{R}_{*P}^2}}+ 
$$ 
$$
+\frac{c_{R}(Q^2)\xi _R(0)}{\pi (Q^2+\mu ^2)\widetilde{R}_{*R}^2}\left( \frac
{Q^2+\mu ^2}{s}\right) ^\eta e^{-\frac{b^2}{\widetilde{R}_{*R}^2}}
$$

where $\xi _{P,R}(0)$ stands for signature factors, \\

$$
\widetilde{R}_{*P}^2=4\alpha _P^{\prime }(0)\ln \frac {s}{Q^2+\mu
^2}+R_{*P}^2(Q^2)\mbox{\ ,}
$$

$$
\widetilde{R}_{*R}^2=4\alpha _R^{\prime }(0)\ln \frac {s}{Q^2+\mu
^2}+R_{*R}^2(Q^2)
$$
are the pomeron and the reggeon "radii". In the framework of Regge approach
the $Q^2$-dependence of radii and residues is neither fixed nor limited.
It gives us a possibility to describe the data in "pure Regge spirit"
(i.e. without any trajectory with the $Q^2$-dependence, or in other words,
without "hard pomerons")

Then
$$
\sigma _{\gamma ^{*}p\rightarrow Vp}(s,Q^2)=4\pi
\int\limits_0^\infty db^2|T_{\gamma ^{*}p\rightarrow Vp}|^2
$$

Figures 3 and 4 show descriptions of the cross-sections
for the processes 
$\gamma ^{*}p\rightarrow \rho _0p$ and $\gamma ^{*}p\rightarrow \phi
p$ and the $Q^2$-dependence of residues and radii needed for that.

One can  obviously conclude:
\begin{enumerate}
\item The data can be described in terms of a generalized Regge-eikonal
mechanism and it   \underline{doesn't demand} introducing new
( $Q^2$-dependent) singularities besides the Regge poles.
\item  The $Q^2$-dependence of residues is rather arbitrary
and does not reveal any contradiction with (quite poor) theory.
\end{enumerate}

Now we shall briefly discuss the $Q^2$-dependence of the cross-section.
Let's analyze the pomeron contribution into the Born term
$$
\hat \delta^{P}_{\gamma^* V} = C^P _{\gamma^* V} (Q^2) e^{(\frac{1}{4} R^2_*(Q^2)+\frac{1}{4}R^2+\alpha ^\prime (0)
log \frac{s}{Q^2+\mu^2})t} \left( \frac{s}{Q^2+\mu^2} \right)^{1+\Delta}
$$
where all powerlike behaviour on $Q^2$ is dictated by absence of actual
(powerlike) scaling violation (i.e. in the leading term) in structure functions of deep inelastic
scattering. The $Q^2$-dependence in $C^P _{\gamma^* V} (Q^2)$ is then supposed to
be weak.

Then we have
$$
\sigma_{\gamma^* V} (s, Q^2)=\left( \frac{1}{Q^2+\mu^2} \right)^{2+2\Delta}
\frac{2s^{2\Delta} |C^P _{\gamma^* V} (Q^2)|^2} {R^2 _* +R^2+ 4\alpha^\prime (0)log[s/(Q^2+\mu^2)]}+...
\mbox{{\hskip  0.5cm}.}
$$
to the extent the Born approximation is valid, the strong $Q^2$-dependence is given by the factor
\begin{equation}
\label{2}
\left( \frac{1}{Q^2+\mu^2}\right)^{2+2\Delta} \mbox{{\hskip 0.2cm}.}
\end{equation}
The experimental data on the $Q^2$-dependence of cross-sections for
exclusive electroproduction of $\rho^0$-mesons with fixed $s$
can be parametrized by a strong dependence like (\ref{2}) 
with the exponent $2.05 \pm 0.09$\cite{1}.
It is curious to compare it with $2+2\Delta \simeq 2.15$. 

Good accordance of our prediction with the data gives us additional optimism.

%----------------Картинка теор.--------------------------------------

\begin{figure}[t]
\label{33}
\vskip  -5cm
\hskip 2cm \vbox to 10cm {\hbox to 10cm{\epsfxsize=10cm\epsffile{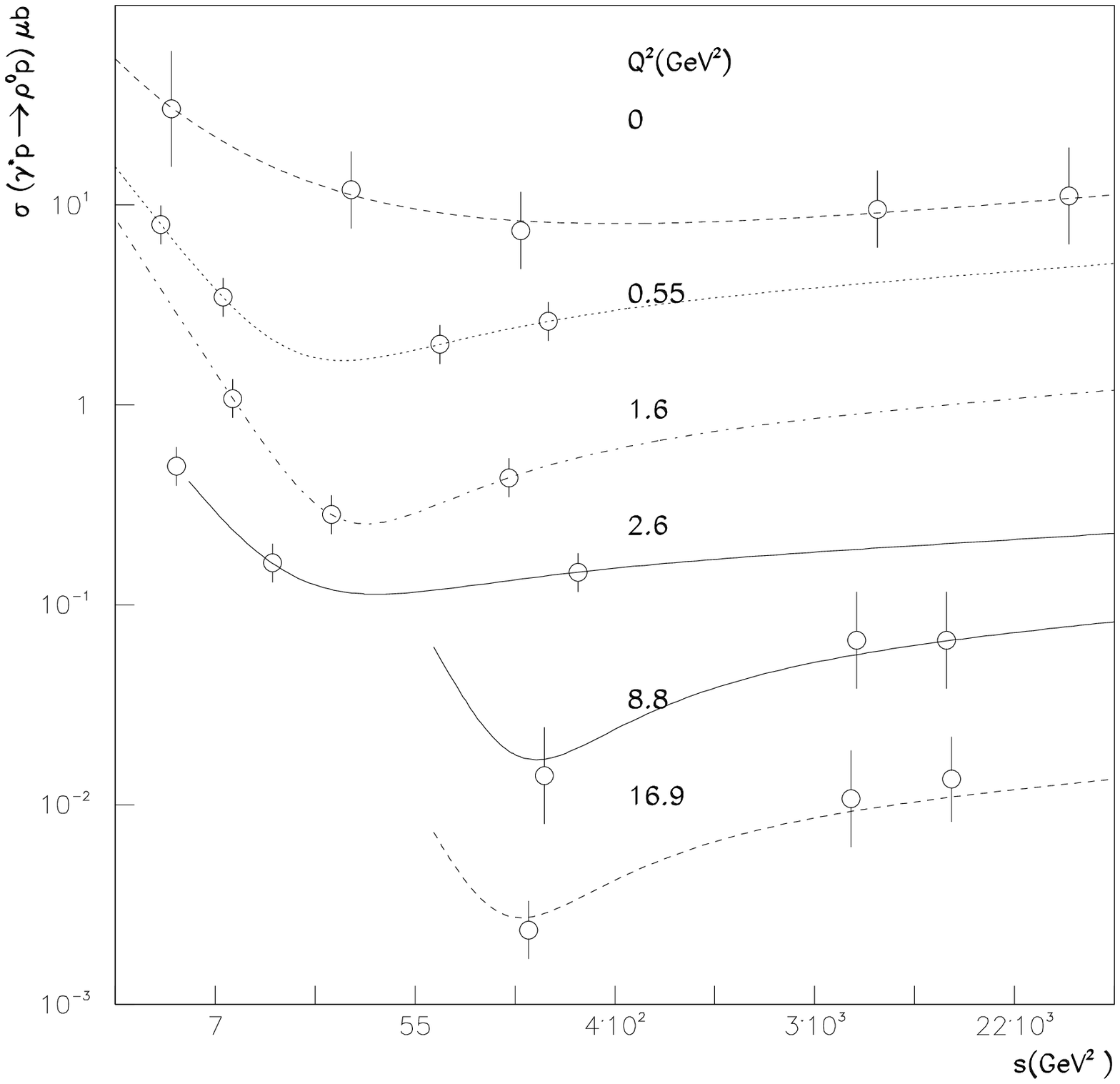}}}
\vskip  2cm
\hskip 2cm \vbox to 10cm {\hbox to 10cm{\epsfxsize=10cm\epsffile{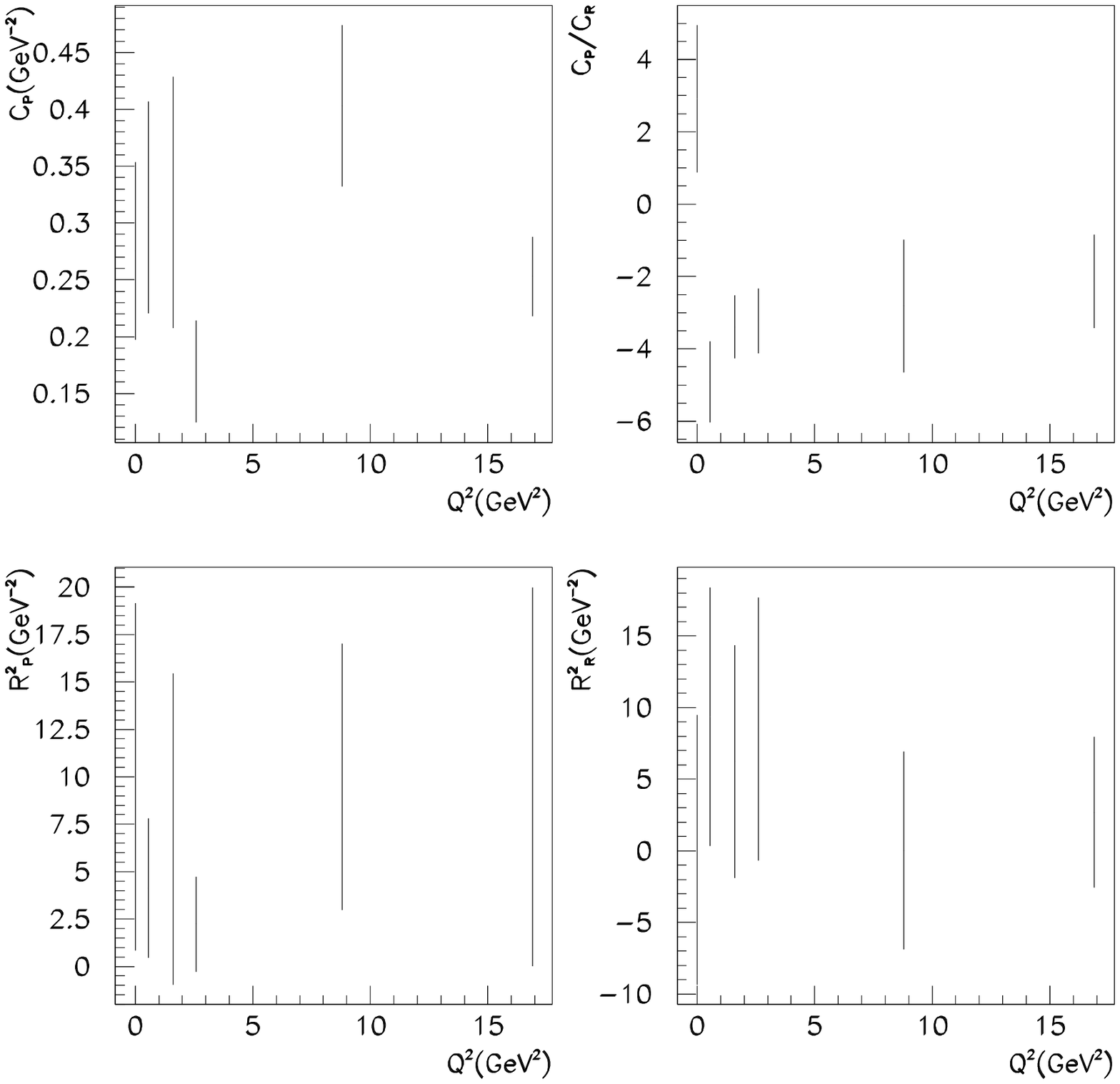}}}
\vskip  2cm
\caption{ The cross-section $ \sigma_ {\gamma p \rightarrow
 \rho_0 p}(s=W^2)$ and the $Q^2$-dependence of parameters.}
%\vskip - 2cm
\end{figure}

\begin{figure}[t]
\label{44}
\vskip -5cm
\hskip 2cm \vbox to 10cm {\hbox to 10cm{\epsfxsize=10cm\epsffile{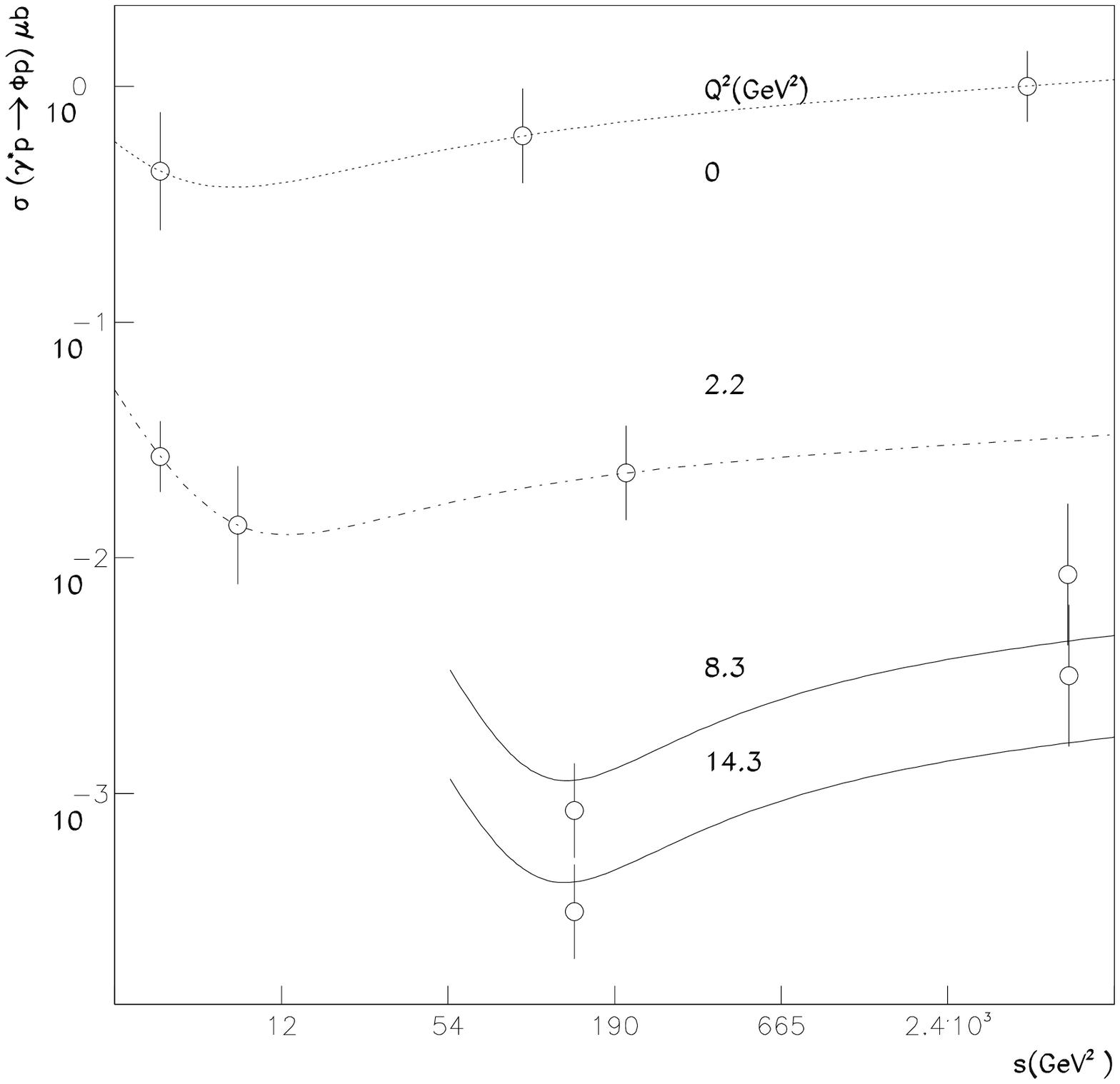}}}
\vskip 2cm
\hskip 2cm \vbox to 10cm {\hbox to 10cm{\epsfxsize=10cm\epsffile{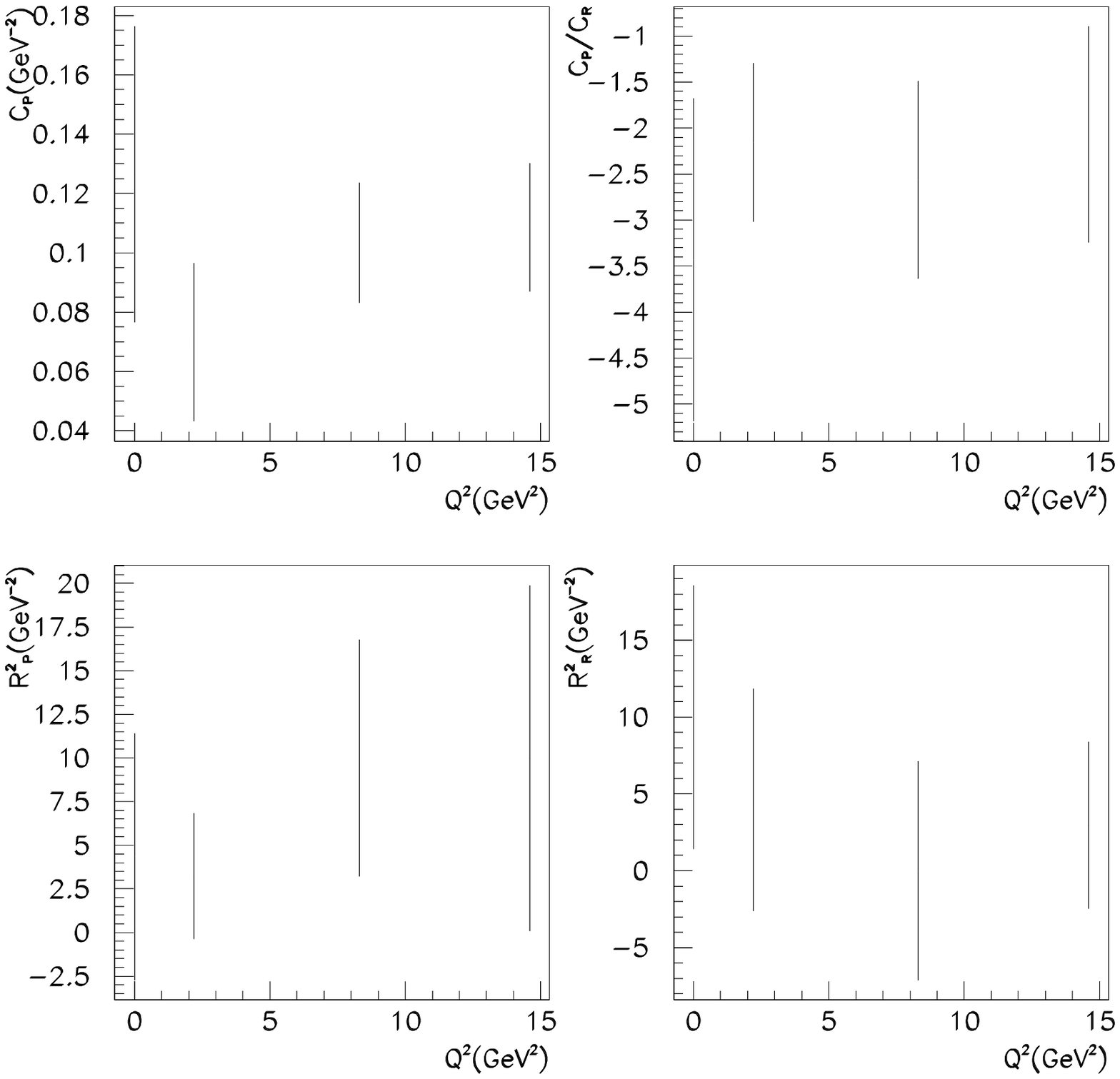}}}
\vskip 2cm
\caption{The cross-section $ \sigma_ {\gamma p \rightarrow
 \phi p}(s=W^2)$ and the $Q^2$-dependence of parameters.}
\end{figure}
%%\newpage

%\begin{figure}[t]
%\label{55}
%%\vskip - 7cm
%\vbox to 14cm {\hbox to 14cm{\epsfxsize=14cm\epsffile{total1.ps}}}
%\caption{Теоретические кривые зависимости $ \sigma^{tot}_ {\gamma^* p \rightarrow
%X}$ от $s=W^2$.}
%\end{figure}

%-----------Заключение-------------------------

\section*{Conclusion}
Thus, in the paper we have managed to establish that
a Regge-eikonal approach generalized for the case of virtual particles 
is valid for the description of the experimental data on exclusive
electroproduction of vector mesons from HERA. The conclusion disproves the
 opinion that new data from HERA
\underline{demand} existence of other, besides Regge poles, singularities in
the $J$-plane. For the sake of justice it is worth  noticing that our work does not
 reject the very possibility
for existence of the non-Regge singularities. Nevertheless, we point out
that general principles of the theory (unitarity in particular)
  play the essential role in answering
the question on determination of relative weight of Regge poles
and possible non-Regge singularities in the complex $J$-plane.

The authors are greateful to E.Predazzi and A.P.Samokhin for useful
discussion, and we also appreciate M.Derrick, J.Whitemore
and A.Marcus for providing us with experimental data and
phenomenological paprametrizations.

\end{document}